\documentclass[conference]{IEEEtran}

\pdfoutput=1

\usepackage{graphics} 
\usepackage{epsfig} 
\usepackage{mathptmx} 
\usepackage{times} 
\usepackage{amsmath} 
\usepackage{amssymb}  
\usepackage{color}
\usepackage{xspace}
\usepackage{cite}
\usepackage{multirow}
\usepackage{mathtools}
\usepackage{relsize}

\allowdisplaybreaks[1]

\newcommand*{\eg}{{e.g.}\@\xspace}
\newcommand*{\ie}{{i.e.}\@\xspace}
\newcommand*{\cf}{{cf.}\@\xspace}
\newcommand*{\wrt}{{w.r.t.}\@\xspace}
\newcommand*{\dof}{{DoF}\@\xspace}
\newcommand*{\figref}{Fig.~\ref}
\newcommand*{\dofs}{{DoFs}\@\xspace}
\newcommand*{\itos}{{I2S}\@\xspace}
\newcommand*{\rom}{{RoM}\@\xspace}
\newcommand*{\wls}{{WLS}\@\xspace}

\newcommand*\iposprocnoise{v^{I^G}_{i,t}}
\newcommand*\ivelprocnoise{v^{\dot I^G}_{i,t}}
\newcommand*\ioriprocnoise{v^{q^{GI}}_{i,t}}

\newcommand*\iangvelmeasnoise{e^{y^\omega}_{i,t}}

\newcommand*\caliborinoise{e^{q^{SI}}_{i,t}}
\newcommand*\calibposnoise{e^{I^{S}}_{i,t}}
\newcommand*\hingenoise{e^H_{k,t}}
\newcommand*\romnoise{e^{\romm}_{k,t}}
\newcommand*\shapeposnoise{e^{I^S_{shape}}_{i}}
\newcommand*\shapeorinoise{e^{q^{SI}_{shape}}_{i}}
\newcommand*\jointvelnoise{e^{J_v}_{k,t}}
\newcommand*\fixedposnoise{e^{S^G_{fix}}_{i,t}}
\newcommand*\constquatnoise{e^{q^{GI}_{0}}_{i,t}}
\newcommand*\constcalibposnoise{e^{q^{SI}_{const}}_{i}}
\newcommand*\constcaliborinoise{e^{I^S_{const}}_{i}}

\newcommand*\ipos{I^G}
\newcommand*\ivel{\dot I^G}
\newcommand*\iori{q^{GI}}

\newcommand*\iangvelmeas{y^\omega}

\newcommand*\calibori{q^{SI}}
\newcommand*\calibpos{I^S}
\newcommand*\hinge{H}
\newcommand*\romm{RoM}
\newcommand*\shapepos{I^S_{shape}}
\newcommand*\shapeori{q^{SI}_{shape}}
\newcommand*\jointvel{J_v}
\newcommand*\fixedpos{S^G_{fix}}
\newcommand*\constquat{q^{GI}_{0}}
\newcommand*\constcalibpos{q^{SI}_{const}}
\newcommand*\constcalibori{I^S_{const}}

\newcommand*\hingeaxis{h^{S}_{i}}

\usepackage{cite}
\usepackage{url}

\usepackage{algorithmic}
\usepackage{algorithm}

\begin{document}
%
\title{\LARGE \bf Towards Self-Calibrating Inertial Body Motion Capture}

\author{\IEEEauthorblockN{Bertram Taetz, Gabriele Bleser, Markus Miezal}
\IEEEauthorblockA{Department of Computer Science \\
University of Kaiserslautern\\
Kaiserslautern, Germany\\
Email: \{taetz,bleser,miezal\}@cs.uni-kl.de}
}


%


\maketitle

\begin{abstract}
This paper presents a novel online capable method for simultaneous estimation
of human motion in terms of segment orientations and positions 
along with sensor-to-segment calibration parameters
from inertial sensors attached to the body.
In order to solve this ill-posed estimation problem, 
state-of-the-art motion, measurement and biomechanical models are combined with new stochastic equations
and priors. These are based on the kinematics of multi-body
systems, anatomical and body shape information, as well as,
parameter properties for regularisation.
This leads to a constrained weighted least squares problem that is solved in a
sliding window fashion.
Magnetometer information is currently only used for initialisation, while the estimation itself works without magnetometers.
The method was tested on simulated, as well as, on real data, captured from a
lower body configuration.
\end{abstract}


%
\IEEEpeerreviewmaketitle

\section{INTRODUCTION}
\label{sec:intro}
\begin{figure}[t]
\centering
\includegraphics[width=\columnwidth]{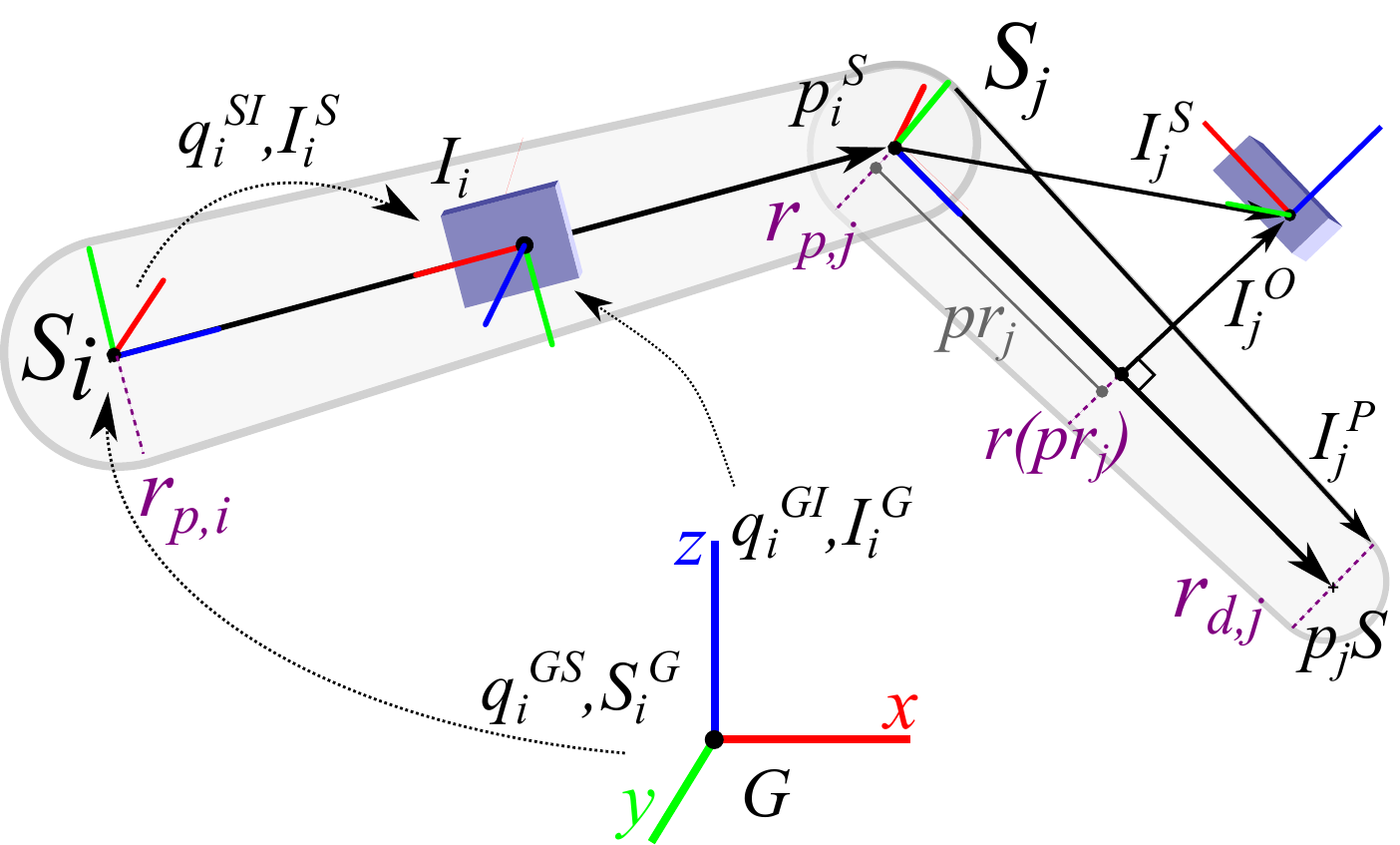}
\caption{Biomechanical model with two segments.}
\label{fig:twosegchain}
\end{figure}

Inertial body motion capture (mocap) has found widespread use in various applications
ranging from robotics \cite{Miller2004} over sports \cite{Ruffaldi2015} and health \cite{Fong2010,Bleser2015a} 
to human-machine-interaction \cite{Bleser2015}.
In particular if in-field assessment of human motion 
is required, body-worn inertial measurement units (IMUs), including 3D accelerometers, gyroscopes and 
often magnetometers, offer key advantages over marker-based optical systems \cite{optitrack},
by not depending on the line of sight or being restricted to laboratory
conditions.
Though mature systems are already available on the
market \cite{Roetenberg2013}, inertial mocap is still subject of research aiming at both increasing the accuracy, 
robustness, as well as, the practicality of such systems. 
One challenge that arises for magnetometer dependent systems is the fact
that man made environments do often not provide a static
magnetic field, thus reduced dependence on magnetometer usage represents an
important aspect of robustness \cite{Ligorio2016a}.
Moreover, in order to deduce the positions and orientations of the body segments comprising 
the biomechanical model via body-worn IMUs, it is crucial to know the 
relative position and orientation of each IMU with respect to (\wrt) the segment
it is attached to (\cf \figref{fig:twosegchain}).
In a kinematic model with rigid segments
and joints, the IMU-to-segment relation (\itos calibration) is typically modelled
as a rigid transformation with six degrees of freedom (\dof)
\cite{Roetenberg2013,Miezal2013,El-Gohary2015,Wenk2015,Kok2014}. 
Providing accurate and easy-to-use calibration mechanisms is therefore an important aspect of practicality.

Though \itos calibration errors, in particular \wrt the orientations, 
immediately lead to errors within the estimated 
segment poses~\cite{DeVries2010,Palermo2014}, 
calibration issues have not been intensively addressed in the
mocap literature, \ie calibration is often assumed to be given,
\eg~\cite{Kok2014,El-Gohary2015}. 
Functional calibration requires the user to precisely perform predefined static poses or movements.
The most simple but also widespread procedures involve only one static pose, 
\eg the so-called N-pose (all segments aligned with the vertical) or T-pose 
(arm segments horizontal), from which all \itos orientations can be determined 
based on measured accelerations and magnetic fields \cite{Roetenberg2013}. 

Note that human anatomy does not allow to precisely perform the N-pose, for
instance, due to the individual's carrying angle, chest and upper arm
circumferences, which all influence the calibration result~\cite{Rettig2009}.
Calibration procedures based on measuring angular velocities during rotations around predefined anatomical axes
result in better consistency with anatomical joint coordinate systems~\cite{DeVries2010}.
However, they usually involve more steps and are less easy to perform by a
subject autonomously \cite{Cutti2008}, \cite{DeVries2010}.
In \cite{Bouvier2015}, different manual alignment and functional calibration methods, including the above mentioned ones, 
are validated against an optoelectronic reference system based on ten healthy subjects instructed by three operators.
The study reports precision in the range $[6,26]^\circ$ 
and reproducibility in the range $[5,10]^\circ$ root mean squared error (RMSE).
It is pointed out that calibration accuracy is more dependent on 
the level of rigor of the experimental procedure (\eg the operator training)
than on the choice of calibration itself.
This underlines the limitation of these methods to require a trained and compliant user,
who, in addition, has the physical and cognitive capabilities to precisely perform the required protocol.

Self-calibration methods determine calibration parameters from sensor measurements 
without prior knowledge or assumptions about the performed movements and 
are therefore particularly interesting when targeting a practical system.
Up to now, such methods appear to be mainly used 
for calibrating \textsl{single sensor units} or packages, in particular when
estimating calibration parameters simultaneously with motion; \eg \cite{Kok2014a}
proposes an offline maximum likelihood estimator for simultaneous IMU
calibration and orientation estimation during arbitrary motion, while
\cite{Wahlstrom2015} proposes an extended Kalman filter based method 
for simultaneous vehicle navigation and smartphone-to-vehicle alignment.
In the recent robotics literature, several examples of 
online \cite{Li2014a} and offline \cite{Cucci2014,Birbach2014} estimation of calibration parameters 
simultaneously with motion can be found.
However, self-calibration methods for human motion tracking appear to be rare. 
In \cite{Seel2014}, offline least squares estimators are proposed for estimating 
calibration parameters of two linked segments with attached IMUs from their measured angular velocities and accelerations.
Parameters include the rotation axis of a revolute joint (\eg the knee) and the 
position of a ball-and-socket joint, given in the reference frames of the two IMUs.
In \cite{Salehi2015} the IMU position calibration method of \cite{Seel2014} 
is initially analyzed \wrt observability and additional constraints from a three-link-segment are
introduced in order to better constrain the optimisation problem. 

Removing the need for precisely executing predefined calibration poses or 
movements, in particular when IMUs have slipped unintentionally during recording, 
is a key requirement for obtaining a truly practical system,
which can be operated by a wide range of non-expert users.
The present work makes a first step into this direction by showing results 
from a novel online-capable, self-calibrating inertial body mocap system. 
The approach is inspired by the offline inertial mocap method
of~\cite{Kok2014} that obtains a constrained weighted least squares (\wls)
estimate for a complete movement from a batch of inertial measurements.
In this paper, a sliding window based constrained \wls
method is proposed for simultaneously estimating the body motion along with the
\itos calibration parameters. The approach can also be used as a moving
horizon approach, thus avoiding a delay in processing streamed data \cite{Rao2001}.
In order to achieve convergence from a wide range of initialisations, 
new stochastic equations and priors are introduced into the objective function.
These are based on multi-body kinematic equations~\cite{Wenk2015}, 
anatomical information (restricted joints and range of motion),
similar to \cite{El-Gohary2015}, and novel inclusion of body shape information
as well as regularisation. 
The convergence behaviour, precision and repeatability of the proposed method were initially tested within both, 
a simulation study and a real data case study, where a subject performed squat exercises. 
On simulated data from a two segment model, reliable convergence with sub-degree precision was observed 
when initialising the \itos orientations with angular offsets up to $95^\circ$.
On real data captured from a lower body with four segments,
repeatable results (below $2^\circ$ difference) were obtained from three different initialisations 
of the sensor-to-segment calibrations including a maximum initial angular
difference of $113^\circ$.
In the following, Section \ref{sec:notation} introduces the notation,
while Sections \ref{sec:statespace} through \ref{sec:opti-method} explain the
proposed method, which is then evaluated in Sections \ref{sec:sim_study} and
\ref{sec:real_study}.
Finally, Section~\ref{sec:conclusion} draws conclusions.

\section{Notation and Biomechanical Model \label{sec:notation}}
As illustrated in \figref{fig:twosegchain}, the human body is represented as a
set of rigid segments $\mathbb{S}$ that are connected through joints,
$\mathbb{J}$.
For each joint, $J_k \in \mathbb{J}$, the connecting segments are collected in $\mathbb{S}_{J_k}$.
A subset $\mathbb{H} \subset \mathbb{J}$ is assumed to be hinge joints 
(\eg the knee has one major rotation axis, which is in general not aligned with
any of the local segment frame axes \cite{Seel2014}) with limited range of motion (\rom),
while the other joints are modelled as ball-and-socket joints.
Inspired by the Hanavan model \cite{Zatsiorsky2002}, each segment $S_i \in \mathbb{S}$ 
is surrounded by a capsule $C_i \in \mathbb{C}$, which represents the soft tissue.

A set of IMUs $\mathbb{I}$ is attached to the body, 
where $I_i \in \mathbb{I}$ is assumed to be on segment $S_i$, 
sitting on the surface of $C_i$.
Each segment $S_i \in \mathbb{S}$ has a local frame attached to it, 
with the origin in the centre of rotation of the proximal joint and the $z$-axis pointing along the segment.
The segment pose \wrt the global frame $G$
is parametrized through an orientation quaternion $q^{GS}_i$ 
and a position $S_i^G$. The segment vector, $p^{S}_i$, 
points along the segment from the proximal to the distal joint or endpoint, 
where $\|p^{S}_i\|$ corresponds to the segment length.
The surrounding capsule $C_i$ has radii $r_{p,i}$ and $r_{d,i}$ 
at the proximal and distal joint, respectively. Note that $r_{d,i}$ is
equivalent with $r_{p,i}$ of the subsequent segment in distal direction.
 Each IMU $I_i$ has a local frame attached to it, 
with origin in the centre of the accelerometer triad, 
coordinate axes aligned to the casing and local $z$-axis pointing orthogonally away from the bottom plate.
The IMU's orientation and position \wrt to $G$ are denoted $q^{GI}_i$ and $I_i^G$, respectively.
The \itos calibrations are defined as the 
relative orientations $q^{SI}_i$ and positions $I_i^{S}$, $\forall I_i \in \mathbb{I}$.

The knowledge required by the proposed method comprises segment lengths, 
capsule radii, hinge rotation axes and ranges of motion.
These can be obtained based on measurements in combination with anthropometric databases \cite{Trieb2013,Zatsiorsky2002}, 
body scanning technologies \cite{Peyer2015,Wasenmueller2015} or calibration
\cite{Seel2014} and have to be determined once per person.

In the following, unit quaternion $q$ and respective rotation matrix $R$ are used interchangeably.

\section{Variables, motion and measurement models}
\label{sec:statespace}
The variables to be estimated include:
\begin{itemize}
\item IMU poses and velocities for each time step $t$: 
$I^{G}_{i,t}, \dot I^{G}_{i,t} q^{GI}_{i,t}, \omega^{GI}_{I,i,t}, \forall I_i \in \mathbb{I}$
\item Segment poses for each time step $t$: $S^{G}_{i,t}, q^{GS}_{i,t}, \forall S_i \in \mathbb{S}$
\item \itos calibrations: $I_i^{S}, q^{SI}_i, \forall I_i \in \mathbb{I}$. 
\end{itemize}
Note, the general approach of defining a redundant variable set 
in combination with constraints, 
rather than relying on a minimal parametrisation (\eg \cite{Miezal2013},
\cite{Miezal2014}) is adopted from \cite{Kok2014}, arguing that this better
accounts for model violations typically appearing in human motion tracking, (\eg due to soft tissue artefacts and anatomical variability).

Note, gyroscope and accelerometer bias models are not in the focus of this paper, but 
can be easily added, see \eg \cite{Kok2014,Bleser2009}.

The motion of each IMU $I_i \in \mathbb{I}$ from time step $t$ to $t+1$, with sampling time $T$, 
is modelled by taking the measured acceleration, $y^a_{i,t}$, as input \cite{Bleser2009},
yielding $\forall I_i \in \mathbb{I}$:
\begin{subequations}
\begin{align}
 I^{G}_{i,t+1} &= I^{G}_{i,t} + T \dot I^{G}_{i,t} + \frac{T^2}{2} R^{GI}_{i,t} ( y^a_{i,t} - \iposprocnoise) + 
\frac{T^2}{2} g^G,\\
\dot I^{G}_{i,t+1} &= \dot I^{G}_{i,t} + T R^{GI}_{i,t} \left(y^a_{i,t} -
\ivelprocnoise\right) + T g^G,\\
q^{GI}_{i,t+1} &= q^{GI}_{i,t} \odot \exp\left(\frac{T}{2}\omega^{GI}_{I,i,t} +
\ioriprocnoise\right),
\end{align}
\label{eq:motion_model}
\end{subequations}
with process noises $v^X_{i,t} \sim \mathcal{N}(0, \Sigma^{X}),\, X \in
\{\ipos,\ivel,\iori\}$.
Here, $g^G$ denotes the global gravity vector, $\odot$ and $\exp$ are the
quaternion product and exponential, respectively. 

For all $I_{i} \in \mathbb{I}$ 
the measured angular velocity, $y^\omega_{i,t}$, is related to the estimated angular velocity via \cite{Bleser2009}:
\begin{equation}
y^\omega_{i,t} = \omega^{GI}_{I,i,t} + \iangvelmeasnoise,\quad \iangvelmeasnoise
\sim \mathcal{N}(0, \Sigma^{\iangvelmeas}).
\label{eq:gyro_meas}
\end{equation}
The reason for modelling the measured accelerations as inputs and the angular velocities as measurements is that
both angular and linear velocities are required as estimation variables (\cf Section \ref{sec:velconstraint}).

\section{Biomechanical model and priors}
\label{sec:constraints}
In the following, the biomechanical model is formulated as constraints,
stochastic equations and priors for the final WLS minimisation. Here, priors correspond 
to equations that include the \itos calibration parameters only.

\subsection{Connected segments constraint and \itos calibration}
These relations are adapted from \cite{Kok2014} and are included as follows.
$\forall J_k \in \mathbb{J}$ with $S_{i}, S_{j} \in \mathbb{S}_{J_k}$:
\begin{equation}
c(S_{j,t}^G, S_{i,t}^G, q^{GS}_{i,t}) = S_{j,t}^G - \left(S_{i,t}^G + R^{GS}_{i,t}
(p^{S}_{i})\right).
\label{eq:connected_segments}
\end{equation}
This constrains the body segments to be attached at the joints.
Moreover, $\forall I_i \in \mathbb{I}$:
\begin{subequations}
\begin{align}
q^{GI}_{i,t} &= q^{GS}_{i,t} \odot q^{SI}_{i,t} \odot \exp\left(\frac{1}{2} \caliborinoise\right), 
\caliborinoise \sim \mathcal{N}(0, \Sigma^{\calibori}),\\
I_{i,t}^{G} &= S_{i,t}^{G} + R^{GS}_{i,t} \left(I^{S}_{i,t} +
\calibposnoise\right), \calibposnoise \sim \mathcal{N}(0, \Sigma^{\calibpos}),
\end{align}
\label{eq:imu_placement}
\end{subequations}
models the fact that IMU and segment poses are coupled through the \itos calibrations,
up to some uncertainty that might compensate for soft-tissue
artefacts \cite{Kok2014}.

\subsection{Velocity at joints}
\label{sec:velconstraint}
In \cite{Wenk2015}, a measurement model for minimising
the linear velocity difference at a joint was proposed in a filter framework. 
The goal was to reduce the dependence on magnetometer usage for heading
drift correction.
In this work we found that this minimisation aids the calibration
estimation (\cf \figref{fig:constraints_comparison}).
The equation can be adapted as follows. $\forall J_k \in \mathbb{J}$ with
$S_{i}, S_{j} \in \mathbb{S}_{J_k}$, $I_i, I_j \in \mathbb{I}$:
\begin{multline}
\jointvelnoise = \dot I_{i,t}^{I} + \omega_{I,i}^{GI} \times
\left(R^{IS}_i (p_i^{S_i} - I_i^{S_{i}})\right) \\
- R^{IG}_i R^{GI}_j\left(\dot I_{j,t}^{I} + (R^{IS}_jI_j^{S}) \times
\omega_{I,j}^{GI}\right),\quad \jointvelnoise \sim \mathcal{N}(0,
\Sigma^{\jointvel}).
\label{eq:velocity_constraint}
\end{multline}

\subsection{Hinge joint and \rom limit} 
Minimising $\forall J_k \in \mathbb{H}$ with $S_{i}, S_{j} \in \mathbb{S}_{J_k}$:
\begin{subequations}
\begin{equation}
\hingenoise = \hingeaxis - R^{SG}_{j,t} R^{GS}_{i,t} \hingeaxis, \quad
\hingenoise \sim \mathcal{N}(0, \Sigma^{\hinge}),
\label{eq:hinge_constraint}
\end{equation}
constrains the rotation of joint $J_k$ to mainly appear around axis $\hingeaxis$,
which is assumed known (\cf Section \ref{sec:notation}). 
Here, the additive noise term $\hingenoise$ accounts for both, an error in the rotation axis and
small rotations around other axes. 
Moreover, rotation angles around this axis outside a predefined range, 
$\text{\rom} \in [\theta_{min,k},\theta_{max,k}]$, are penalised by minimising:
\begin{equation}
\romnoise = 
\begin{cases}
\theta_{min,k}-\theta_{k,t} & \text{for } \theta_{k,t} < \theta_{min,k}\\
\theta_{k,t} - \theta_{max,k} & \text{for } \theta_{k,t} > \theta_{max,k}\\
0 & \text{otherwise },
\end{cases}
\end{equation}
\label{eq:hinge}
\end{subequations}
with $\theta_{k,t} = 2 \arccos([q^{SG}_{j,t} \odot q^{GS}_{i,t}]_w)$ 
and $\romnoise \sim \mathcal{N}(0, \Sigma^{\romm})$.

\subsection{Body shape prior}
Assuming each IMU to be mounted approximately on the surface of the respective
body segment, approximated via a capsule (\cf \figref{fig:twosegchain}), 
is a reasonable assumption that can be used to guide both the \itos orientation and position estimation.
As illustrated in \figref{fig:twosegchain}, 
let $pr_{i,t}$ be the length of the orthogonal projection of IMU position $I^S_{i}$ 
onto the segment $p^{S}_{i}$, and let $I^O_{i}$ be the vector part of $I^{S}_i$ 
which is orthogonal to the segment.
Minimising, $\forall I_i \in \mathbb{I}$:
\begin{subequations}
\begin{equation}
\shapeposnoise = 
\begin{cases} 
I^{S}_{i} - r_{p,i} \overline{I^{S}_{i}} & \text{for } pr_{i} < 0 \\
I^{O}_{i} - r(pr_{i}) \overline{I^O_{i}} & \text{for } 0 \leq pr_{i} \leq
\|p^{S_i}\|\\
(I^{S}_{i} - p^{S}_i) - r_{d,i} \overline{I^{S}_{i} - p^{S}_i} & \text{for }
pr_{i} > \|p^{S}_i\|, \end{cases}
\label{equ:shape_pos}
\end{equation}
with $\overline{v} \coloneqq \frac{v}{\|v\|}$, 
$r(pr_{i}) \coloneqq r_{p,i} + \frac{pr_{i}}{\|p^{S}_i\|}(r_{d,i} - r_{p,i})$
and $\shapeposnoise \sim \mathcal{N}(0, \Sigma^{\shapepos})$
allows $I^S_{i,t}$ to move on the surface of the capsule while penalising orthogonal displacements.
The above model can be extended to other shapes by modifying the computation of the
radius $r(pr_{i})$. Even more complex shapes (\eg from a body scanner) could be included, 
given that the complete surface and outer normals of the shape are known.
Moreover, as illustrated in \figref{fig:twosegchain}, let $I^P_{i}$ be a
vector parallel to the capsule surface. Minimising:
\begin{equation}
\shapeorinoise = [R^{SI}_{i}]_{3,1:3} \overline{I_{i}^P}, \quad
\shapeorinoise \sim \mathcal{N}(0, \Sigma^{\shapeori}),
\label{equ:shape_or}
\end{equation}
\label{eq:shape}
\end{subequations}
 constrains the IMU's
$z$-axis to point orthogonally to the capsule surface. Here, $R_{3,1:3}$ picks
the third row of the rotation matrix $R$.
Without loss of generality, we assume here that the $z$-axis of the IMU points 
away from the surface it is mounted on.
The positive effect of this prior can be observed in \figref{fig:constraints_comparison}.
Note, this prior only depends on the calibration parameters and is therefore not time dependent.

\subsection{Fixed segment position}
In some movement scenarios, it is known that some segments are stationary at specific points,
\eg that the soles of the feet stay on the ground. 
Let $\mathbb{F} \subset \mathbb{S}$ define a subset of segments with fixed position.
Minimising, $\forall S_i \in \mathbb{F}$:
\begin{equation}
\fixedposnoise = p^G_{i,fix} - \left(S^G_{i,t} + R^{GS}_{i,t} p^S_{i,fix}\right), 
\fixedposnoise \sim \mathcal{N}(0, \Sigma^{\fixedpos})
\label{eq:fixed_position}
\end{equation}
constrains the fixed position $p^S_{i,fix}$ to coincide with the known global position 
$p^G_{i,fix}$ at each time step $t$.
Alternative methods to reduce drift are, for instance, to enforce zero velocity
based on detections \cite{Skog2010} or to enforce a known mean acceleration \cite{Kok2014}.

\section{Sliding window based optimisation \label{sec:opti-method}}
In \cite{Kok2014} it is suggested to calculate an offline maximum a posteriori (MAP) estimate of the IMU and segment poses
of all time steps, by solving a global constrained WLS problem, where \eqref{eq:connected_segments}
enters as hard equality constraints. In order to reduce the computation time,
the optimisation works on accumulated measurement data. Moreover, the \itos calibrations are assumed known.
In order to obtain an online-capable approach a sliding window based
\wls estimate is proposed in this work. It sequentially processes overlapping
batches of few IMU data as they are captured, in order to estimate both kinematics and
\itos calibrations. 
Let $\{y^a, y^\omega\}_{t=0:w-1}^b, w \geq 2$ 
be the sequence of IMU data available in batch $b \geq 0$, with $y_t \coloneqq [y^a, y^\omega]_t^T$.
Moreover, for $b>0$, let $\{y^a, y^\omega\}_{w-1}^{b-1} = \{y^a, y^\omega\}^{b}_0$,
\ie let the last time step of batch $b-1$ correspond to the first time step in batch $b$,
so that each new batch contains $w-1$ new time steps and the size of each batch is $w$.
Then, the variables to be estimated for batch $b$ comprise (\cf Section \ref{sec:statespace}):
time-varying IMU kinematics and segment poses:
\begin{subequations}
\begin{align}
x^b &= 
\left\lbrace \left\lbrace I^{G}_{i,t}, \dot I^{G}_{i,t} q^{GI}_{i,t}, \omega^{GI}_{I,i,t}\right\rbrace_{I_i \in \mathbb{I}}, 
\left\lbrace S^{G}_{i,t}, q^{GS}_{i,t}\right\rbrace_{S_i \in \mathbb{S}} \right\rbrace_{t=0:w-1}^b
\intertext{and static \itos calibrations}
z^b &= \left\lbrace I_i^{S}, q^{SI}_i\right\rbrace_{I_i \in \mathbb{I}}^b.
\end{align}
\label{eq:batch_state}
\end{subequations}
Overall this gives the following dimensionality for the state per batch $[x^b,
z^b]^T \in \mathbb{R}^{w(13|\mathbb{I}| + 7|\mathbb{S}|) + 7|\mathbb{I}|}$.

\subsection{Batch initialisation and regularisation}
\label{sec:priors}
Measurement noise, model errors and motion that is available in the small
batches can significantly influence the estimation result and cause instant
changes in both the estimates of the time-varying variables at the batch overlap and the
estimated \itos calibrations.
These are $x_{w-1}^{b-1}$, $x_{0}^{b}$ and $z^{b-1}$, $z^{b}$, respectively.
To reduce this effect, different regularisation priors are introduced.
Minimising $\forall I_i \in \mathbb{I}$
\begin{equation}
\constquatnoise = 
\begin{cases} 
2 \log \left(q^{IG,b-1}_{i,w-1} \odot q^{GI,b}_{i,0}\right) & \text{for } b > 0 \\
2 \log \left(q^{IG}_{i,init} \odot q^{GI,b}_{i,0}\right) & \text{else }
\end{cases},
\label{eq:imu_ori_prior}
\end{equation}
with $\constquatnoise \sim \mathcal{N}(0, \Sigma^{\constquat})$
penalises sudden changes of the estimated IMU orientations for the overlap of $b-1$ and $b$.
Note, in a moving horizon context this term corresponds to the arrival cost for
the variables.
Here, $\log$ denotes the quaternion logarithm.
Note, for $b=0$ the initial quaternions $q^{GI}_{i,init}$ are obtained using the TRIAD algorithm \cite{Shuster1981}.
This is currently the only point, where magnetometer data is used in the
proposed method.
Regularising only the initial IMU orientations in each batch turned out to
be sufficient to produce a smooth trajectory.

Moreover, for $b>0$ and $\forall I_i \in \mathbb{I}$, minimising
\begin{subequations}
\begin{align}
\constcalibposnoise &= \log\left(q^{IS}_{i,b-1} \odot q^{SI}_{i,b}\right),\quad  \constcalibposnoise
\sim \mathcal{N}(0, \Sigma^{\constcalibpos})\\
\constcaliborinoise &= I^{S}_{i,b} - I^{S}_{i,b-1}, \quad
\constcaliborinoise \sim \mathcal{N}(0, \Sigma^{\constcalibori})
\end{align}
\label{eq:calib_prior}
\end{subequations}
penalises sudden changes in the \itos calibrations between batch $b-1$ and $b$.
Obviously, the amount of change depends on the noise covariances,
which have to be adapted appropriately in order to enable 
convergence to a stationary or only slightly varying calibration, \ie the correct one.

For this, a convergence indicator has been defined based on the following observations:
First, in the presence of motion, the residuals of the velocity constraint are biased, 
if one or both of the respective \itos calibrations are incorrect.
Second, if an \itos calibration stays rather constant over a history of $h$ batches, 
despite motion and high covariances in \eqref{eq:calib_prior}, a feasible
calibration is indicated.
Hence, if:
\begin{multline}
\frac{1}{w} \frac{1}{|\mathbb{J}|} \left\| \sum_{t=0}^{w-1} \sum_{k \in
\mathbb{J}} \left(e^{J_v}_{k,t}\right)^b \right\|_2 < th^{J_v} \bigwedge \\
\frac{1}{h}\frac{1}{|\mathbb{I}|} \left\| \sum_{l=b-h}^b \sum_{I_i \in
\mathbb{I}} 2\log\left(q^{IS,l-1}_i \odot q^{SI,l}_i\right) \right\|_2 <
th^{q^{SI}} \bigwedge \\
\frac{1}{h}\frac{1}{|\mathbb{I}|} \left\| \sum_{l=b-h}^b \sum_{I_i \in
\mathbb{I}} I^{S,l}_i - I^{S,l-1}_i \right\|_2 < th^{I^{S}},
\label{eq:convergence_detector}
\end{multline}
with $b>h$, convergence is assumed and the covariances 
$\Sigma^{\constcalibori}_t$ and $\Sigma^{\constcalibpos}_t$
are both decreased by a factor $f$ ($h=10$, $th^{J_v}$=0.01, $th^{q^{SI}}$=0.01, $th^{I^{S}}$=0.05, $f=10$ in the experiments).
To derive the thresholds for the above indicators based on a statistical test and 
realize an adaptive covariance update is part of our future work.

\subsection{MAP estimate and resulting \wls problem}
Starting from the MAP estimate for batch $b$ (\cf \cite{Kok2014}):
\begin{multline}
\min_{x^b,z^b} 
\underbrace{- \sum_{t=1}^{w-1} \log
p(x^b_t| x^b_{t-1}, z^b)}_{\text{motion model}} 
\underbrace{- \sum_{t=0}^{w-1} \log
p(y^b_t| x^b_{t}, z^b)}_{\text{measurement and biomechanical models}}\\
\underbrace{-\log p(x^b_0|y^b_0)}_{\text{batch initialisation}}
\underbrace{- \log p(z^b)}_{\text{priors}}\\
s.t. \quad c(x^b)=0,
\label{equ:basic_problem}
\end{multline}
the constrained \wls problem can now be derived by appropriately incorporating
all the above mentioned models, constraints, stochastic equations and priors.
Removing all constant terms, this yields:
\begin{multline}
\min_{x^b,z^b} 
\sum\limits^{w-1}_{t = 1} \sum_{I_i \in \mathbb{I}}
\Bigg(
\underbrace{
  \left\|\iposprocnoise\right\|^2_{(\Sigma^{\ipos})^{-1}}
+ \left\|\ivelprocnoise\right\|^2_{(\Sigma^{\ivel})^{-1}}
+ \left\|\ioriprocnoise\right\|^2_{(\Sigma^{\iori})^{-1}}
}_{\text{motion model \eqref{eq:motion_model}}}
\Bigg) \\
+\sum\limits^{w-1}_{t = 0} \Bigg( \sum_{I_i \in \mathbb{I}}
\Bigg(
\underbrace{
  \left\|\iangvelmeasnoise\right\|^2_{(\Sigma^{\iangvelmeas})^{-1}}
}_{\text{gyr. model \eqref{eq:gyro_meas}}} 
+ \underbrace{
  \left\|\caliborinoise\right\|^2_{(\Sigma^{\calibori})^{-1}}
+ \left\|\calibposnoise\right\|^2_{(\Sigma^{\calibpos})^{-1}}
}_{\text{\itos calibrations \eqref{eq:imu_placement}}}
\Bigg) \\
+ \sum_{J_k \in \mathbb{J}}
\underbrace{
\left\|\jointvelnoise\right\|^2_{(\Sigma^{\jointvel})^{-1}}
}_{\text{vel. at joints \eqref{eq:velocity_constraint}}} 
+ \sum_{J_k \in \mathbb{H}}
\Bigg(
\underbrace{
\left\|\hingenoise\right\|^2_{(\Sigma^{\hinge})^{-1}} +
\left\|\romnoise\right\|^2_{(\Sigma^{\romm})^{-1}}
}_{\text{hinge joints \eqref{eq:hinge}}}
\Bigg) \\
+ \sum_{S_i \in \mathbb{F}}
\underbrace{
  \left\|\fixedposnoise\right\|^2_{(\Sigma^{\fixedpos})^{-1}}
}_{\text{fixed seg. pos. \eqref{eq:fixed_position}}}
\Bigg)
+ \sum_{I_i \in \mathbb{I}}
\Bigg(
\underbrace{
\left\|\constquatnoise\right\|^2_{(\Sigma^{\constquat})^{-1}}
}_{\text{initialisation \eqref{eq:imu_ori_prior}}} \\
+ 
\underbrace{\left\|\constcaliborinoise\right\|^2_{(\Sigma^{\constcalibori})^{-1}}
+ \left\|\constcalibposnoise\right\|^2_{(\Sigma^{\constcalibpos})^{-1}}
}_{\text{if $b>0$, \itos calibration prior \eqref{eq:calib_prior}}} \\
+ \underbrace{
  \left\|\shapeposnoise\right\|^2_{(\Sigma^{\shapepos})^{-1}}
+ \left\|\shapeorinoise\right\|^2_{(\Sigma^{\shapeori})^{-1}}
}_{\text{body shape prior \eqref{equ:shape_pos}, \eqref{equ:shape_or}}}
\Bigg) \\
s.t. \, c(x^b)=0. 
\label{eqn:minimization_problem}
\end{multline}
Note, the terms in \eqref{eqn:minimization_problem} are obtained from the referenced stochastic models and priors by 
reformulating the latter so that the noises are isolated on the left side.
This constrained \wls problem can be solved in different ways: The
hard constraint in \eqref{eqn:minimization_problem} can be enforced using an
infeasible start Gauss-Newton method, as suggested in \cite{Kok2014},
 see \cite{Boyd2004}.
Another possibility is to include this constraint as soft constraint, by adding the term as
stochastic equation with a low covariance matrix.
This leads to an unconstrained \wls problem that can be solved using a standard
solver like Gauss-Newton \cite{Boyd2004} or the Levenberg-Marquardt method
\cite{Levenberg1944}.
An inclusion using general nonlinear optimisation techniques, such as Augmented Lagrangian 
or an inner point method, are also possible, however, these methods usually have
a higher computational cost \cite{GVK502988711}.
\subsection{Initialisation}
\label{sec:init}
For $b=0$, all IMU orientations $\{q^{GI}_{i,t}\}_{I_i \in \mathbb{I}, t=0:w-1}$ are initialised with $q^{GI}_{i,init}$
(\cf Section \ref{sec:priors}),
while all other time-varying variables are initialised with standard values, \ie zero vectors or identity quaternions.
The effect of using different initial values for the \itos calibrations are analyzed in Sections \ref{sec:sim_study}
and \ref{sec:real_study}.

For $b>1$, all time-varying estimation variables $x^b_{t=0:w-1}$
are initialised with $x_{w-1}^{b-1}$. 
This mimics the idea that the current estimate provides a good predictor for the
future, assuming that all the variables change smoothly.
Moreover, the \itos calibrations $z^b$ are also initialised from the previous batch $z^{b-1}$.

\subsection{Tuning parameter settings}
\label{sec:parameters}
All covariance matrices in \eqref{eqn:minimization_problem} 
can be considered tuning parameters for the proposed algorithm. 
However, as already mentioned in \cite{Kok2014}, the algorithm was rather insensitive \wrt
the majority of covariance settings in a large range. 
Therefore, if not otherwise mentioned in the following, the covariances in \eqref{eqn:minimization_problem}
were all set to identity.
Since \eqref{eq:velocity_constraint} was found to be sensitive \wrt noisy measurements, 
the associated covariance was increased to $\Sigma^{\jointvel}=\text{diag}(10,10,10)$. 
The algorithm was found to be most sensitive \wrt the covariances associated 
to the body shape prior \eqref{eq:shape} and the regularisation \eqref{eq:imu_ori_prior}. 
Recall that the regularisation prior influences the amount of variation of the \itos
calibrations between batches and that the capsule model might only be a rough
approximation of the real body shape. In order to allow enough variation
for the \itos calibrations to adjust towards the correct values from batch to batch, as well as, be robust
against shape variations, the associated covariances are
initially set to $\Sigma^{\shapepos}, \Sigma^{\shapeori},
\Sigma^{\constcalibpos}, \Sigma^{\constcalibori} = \text{diag}(100, 100, 100)$.
The batch size is $w=10$ with an overlap of $1$, if not mentioned otherwise in the experiments.


\section{Simulation case study}
\label{sec:sim_study}
One major challenge of the sliding window approach is to enable convergence of the \itos calibration parameters
from a wide range of initialisations, despite the limited information available in each batch.
To analyze the convergence behaviour of the proposed method (including the currently heuristic convergence indicator 
solution \eqref{eq:convergence_detector}) was the focus of the simulation study.
Note, in this study, the main focus was on the \itos and segment orientation
errors, since accurate tracking of segment orientations is of primary
interest for our applications \cite{Bleser2015a}.

Therefore, IMU data was simulated from $2$ IMUs $\mathbb{I} = \{I_{0},I_1\}$,
mounted on a biomechanical model with $2$ segments $\mathbb{S} = \{S_{0},S_1\}$
and capsules $\mathbb{C} = \{C_{0},C_1\}$ 
($||p^S_i||=0.3 m, r_{p,i}=r_{d,i}=0.1 m, i=0,1$), as well as, two joints
$\mathbb{J} = \{J_0,J_1\}$, with $J_0$ being a ball-and-socket joint and $J_1 \in \mathbb{H}$ being a hinge joint 
($h^S_1 = [1, 0, 0]^T$, \rom $=[\theta_{min,1}=0,\theta_{max,1}=162]^\circ$).
The target \itos calibrations were chosen as indicated in \figref{fig:twosegchain}.

For animating the biomechanical model, $S^G_0$ was kept stationary in the origin,
\ie with $p^S_{0,fix} = p^G_{0,fix} = [0,0,0]^T$, and an angle sequence $\{\{\phi\}_{d=0:3}\}_{t=0:628}$ 
was generated for each rotational \dof $d$ (\ie $d_{0:2}$ for $J_0$ and $d_3$ for $J_1$) using:
\begin{equation}
\phi(\alpha(t))_d = \sin\left( \frac{\alpha(t)}{2} \right) \sin(\alpha(t)) \pi,
\end{equation}
with $\alpha(t) = \frac{2\pi}{629} t$. Here, $\phi(\alpha(t))_3$ was clipped to $\theta_{max,1}$
in order to respect the \rom. 
This provided sufficient variations in all \dofs, with smoothly varying and
periodically increasing and decreasing angular velocities, as well as, direction changes.

Now, for each $I_i \in \mathbb{I}$, $441$ starting values $\{q^{SI}_{i,l}, I^S_{i,l}\}_{l=0:440}$
were generated, yielding $882$ tests altogether. This was done by systematically sampling rotation offset tuples 
$(\beta,\gamma)_l \in \bigcup_{\beta, \gamma \in R}$ with $R=\{-100^\circ,
-90^\circ,\ldots, 90^\circ, 100^\circ\}$ and applying those to the target calibration $q^{SI}_{i},
I^S_i$ using:
\begin{subequations}
\begin{align}
q^{SI}_{i,l} & = q_{z}(\gamma_l) \odot q^{SI}_{i} \odot q_{z}(\beta_l)\\
I^S_{i,l} &= R_{z}(\gamma_l) I^S_{i}.
\label{eq:simulated_calibrations}
\end{align}
\end{subequations}
Here, $q_{z}(\gamma_l)$ denotes a rotation of angle $\gamma_l$ around the segment's $z$-axis, 
where $I^S_i$ is moved accordingly on the capsule surface, and 
$q_{z}(\beta_l)$ is a rotation of angle $\beta_l$ around the IMU's $z$-axis.

Note, when calculating the absolute angular offset between initial and target orientation using:
\begin{align}
q^{SI}_{i,l} \angle q^{SI}_{i} \coloneqq \left|2 \arccos \left[q^{SI}_{i,l} \odot q^{IS}_{i}\right]_0\right|,
\label{eq:ang_offset}
\end{align}
the above variations include initial angular offsets up to $131.19^\circ$
and position offsets up to $0.15m$. 

\subsection{Convergence behaviour}
\begin{figure}
\centering
\includegraphics[width=\columnwidth]{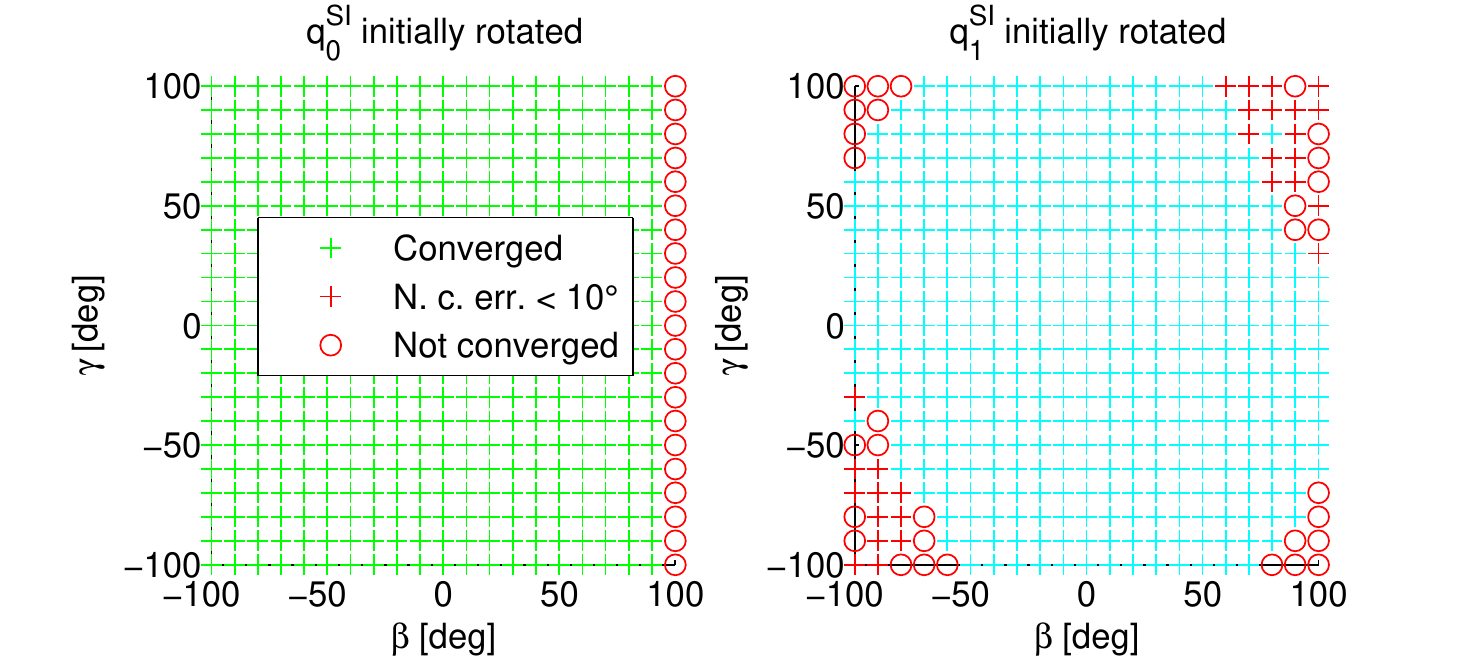}
\caption{Simulation study: convergence behaviour ($+$: true positive (convergence detected and actually converged), 
$\circ$: true negative, red $+$: false negative).}
\label{fig:sim_convergence}
\end{figure}
\begin{figure}
\centering
\includegraphics[width=\columnwidth]{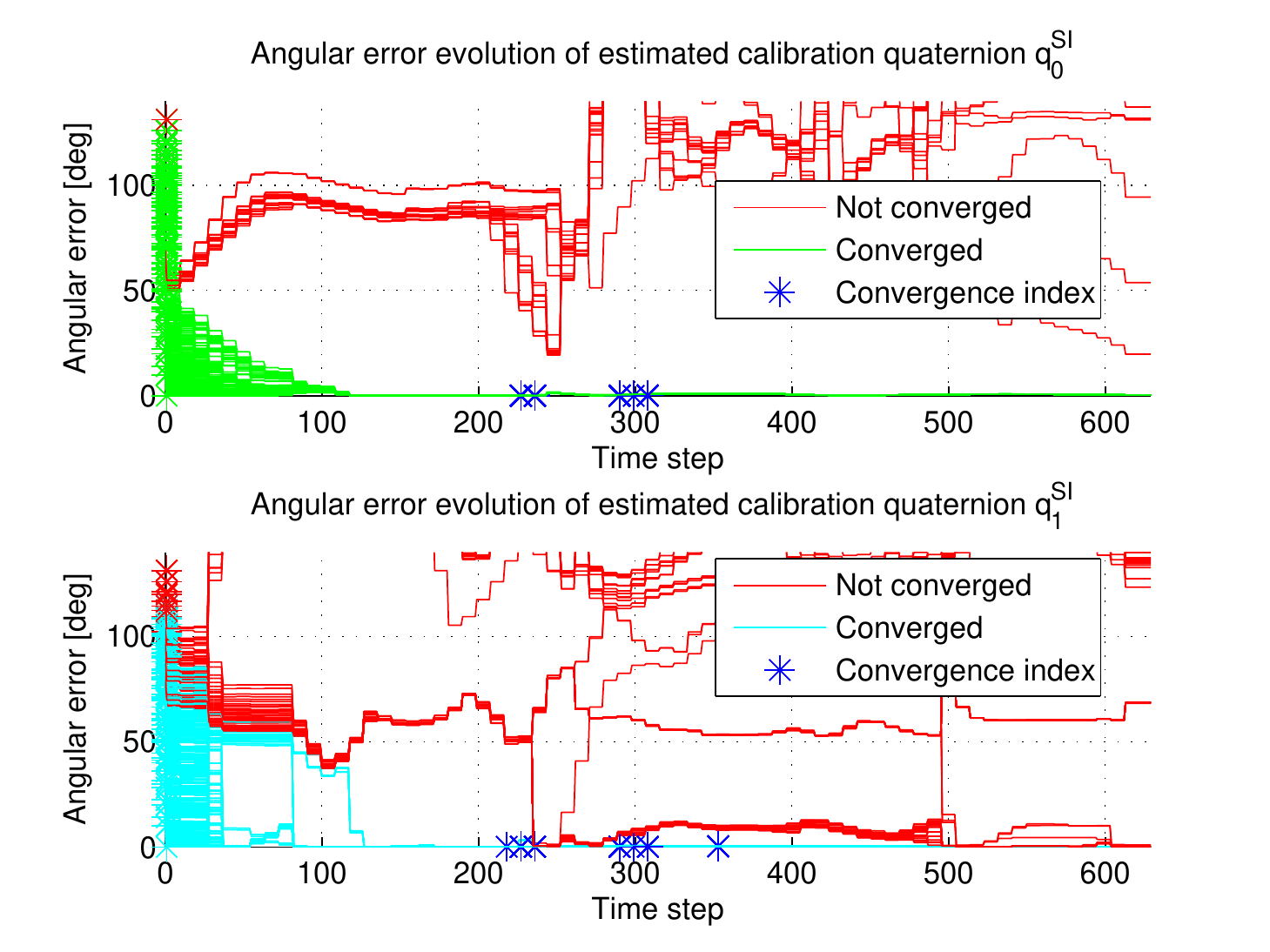}
\caption{Simulation study: angular error evolutions. 
Note, in all plots, a star at time step $0$ marks the initial angular offset as calculated using Equation \eqref{eq:ang_offset}. Moreover, the two graphs show the results for the IMU, which was initialised incorrectly.}
\label{fig:sim_error_evolution}
\end{figure}
\figref{fig:sim_convergence} provides an overview of the convergence
behaviour.
All initial \itos calibration-offsets are marked \wrt both,
whether convergence was detected by the method or not (Equation \eqref{eq:convergence_detector})
and whether this detection could be considered correct or not (based on a threshold of $10^\circ$ mean angular error).
\figref{fig:sim_error_evolution} provides more detailed information about
the actual angular error evolution, as well as, the time, where convergence was detected.

For the tests, where $I_0$ was initialised incorrectly, 
$420$ out of $441$ tests ($=95.24\%$) converged correctly within time step range $[227,308]$.
The minimal offset angle (\cf \eqref{eq:ang_offset}), where the test did not converge, was $100.00^\circ$.
The maximum angle, where convergence was correctly detected, was $131.19^\circ$.
For the tests, where $I_1$ was initialised incorrectly, $382$ out of $441$ tests ($=86.62\%$) converged correctly within time step range $[218,353]$.
Moreover, for $28$ tests, there was no convergence detected,
while both estimated \itos orientations showed a mean angular error below $10^\circ$ 
(mean taken over all time steps after $308$ for $I_0$ and $353$ for $I_1$).
This could be interpreted as false negatives, considering the error ranges of established calibration methods as 
mention in Section \ref{sec:intro}.
In \figref{fig:sim_error_evolution}, these tests appear as red plots with a rather low angular error.
The minimal offset angle, where the test did not converge, was $96.72^\circ$.
The maximum angle, where convergence was detected, was $116.46^\circ$.

\begin{table}%
\centering
\caption{Simulation study: error statistics for all converged tests.}
\begin{tabular}{|l|c|c|}
\hline
				                  & $\mathbf{I_0}$: mean, std, max & $\mathbf{I_1}$: mean, std, max \\\hline
Pos. error $\mathbf{I^S} [m]$        & $0.013$, $0.014$, $0.053$ & $0.008$, $0.004$, $0.024$ \\\hline
Abs. ang. error $\mathbf{q^{SI}} [\circ]$ & $0.574$, $0.358$, $3.435$ & $0.136$, $0.119$, $0.629$ \\\hline
Abs. ang. error $\mathbf{q^{SG}} [\circ]$ & $0.575$, $0.357$, $3.442$ & $0.136$,
$0.119$, $0.630$ \\\hline
\end{tabular}
\label{tab:simulation_results}
\end{table}
Table \ref{tab:simulation_results} provides error statistics for the estimated \itos calibrations,
as well as, the segment orientations, computed from all converged tests, after the time, where convergence was detected.
The average calibration errors are small, in sub-degree range for both IMUs (though comparably higher for $I_0$), 
in the order of a centimetre for $I_0$ and in the order of millimetres for $I_1$, 
providing overall good precision. 
Moreover, low standard deviations and maximum values indicate good repeatability, 
though the maximum values for $I_0$ are higher than those for $I_1$. 
Note, for $q^{SI}_0$, all angular errors above $2^\circ$
were observed with initial angular offsets of $q^{SI}_1$ above $90^\circ$,
indicating a stronger error propagation from $I_1$ to $I_0$.
Another interesting observation is that the mean angular errors of $q^{SI}$
are nearly identical to those of $q^{SG}$.
This indicates that (a) orientation calibration errors propagate linearly into the estimated segment 
orientations and (b) that the tracking does not add significant errors given the otherwise perfect conditions in this simulation study. 

In summary, the proposed method correctly converged from a wide range of \itos initialisations, 
up to above $95^\circ$ of angular offset for both IMUs.
There were no false positive convergence detections. 
For $I_1$, there were a few false negatives.

\subsection{Contribution of different model equations (case study)}
\begin{figure}
\centering
\includegraphics[width=\columnwidth]{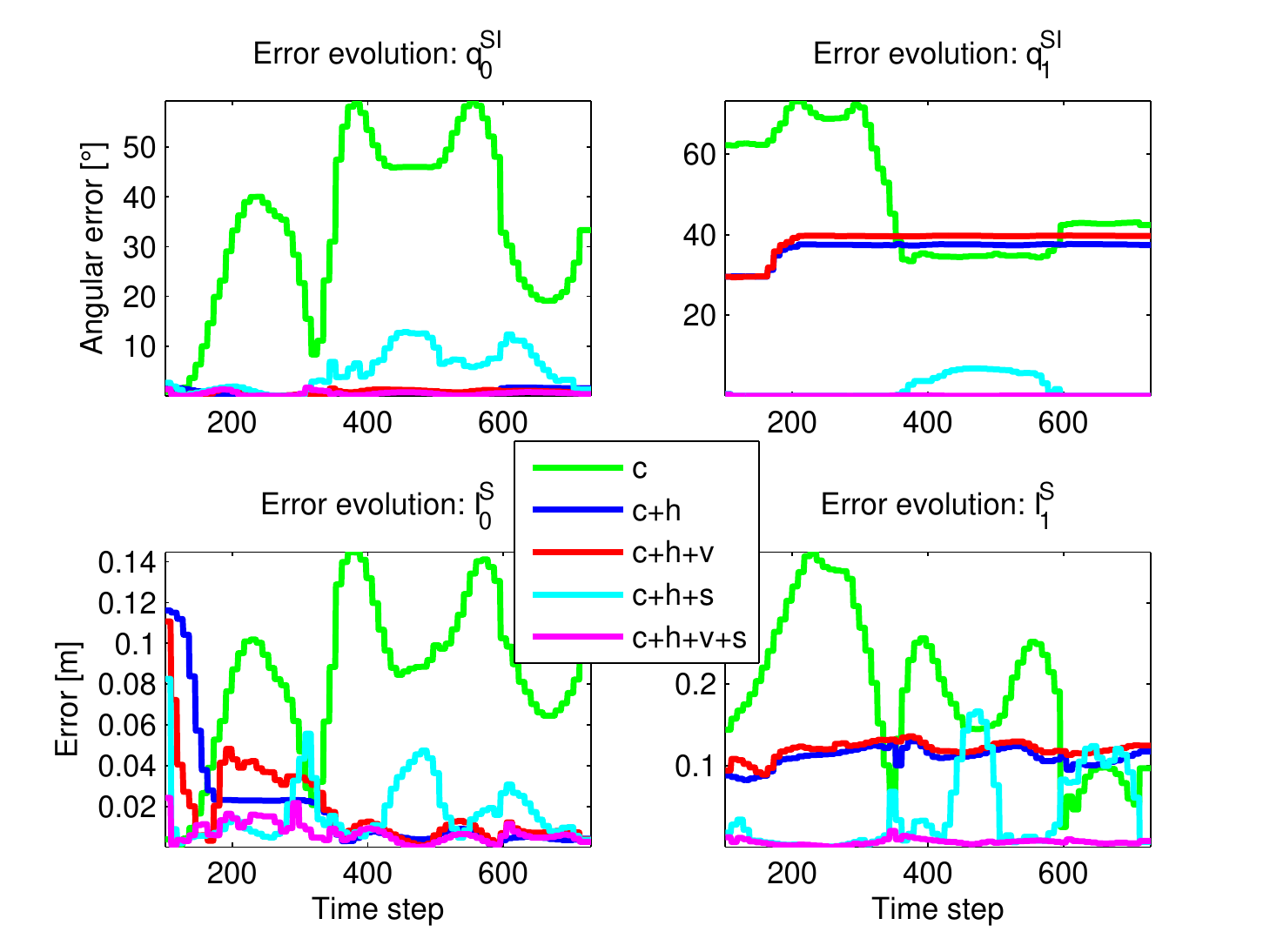}
\caption{Simulation case study: calibration error evolutions when using different combinations of constraints,
stochastic equations and priors
(c=connected segments, h=hinge, v=velocity, s=shape).}
\label{fig:constraints_comparison}
\end{figure}
As major contribution of this work we consider the combination of different 
constraints, stochastic equations and priors in order to sufficiently constrain the 
estimation problem, so that the \itos calibrations can be correctly estimated under motion
from only small batches of data.
\figref{fig:constraints_comparison} shows the contributions of the different constraints and priors
exemplified for one representative simulation test ($\gamma=45^\circ, \beta=-45^\circ$ applied to $I_1$).
Here, it is clearly visible that only the combination of all proposed constraints and priors
leads to convergence of the estimated \itos calibrations of both segments.
A more in depth study under different movements and configurations is planned as future work.

\section{Real data case study}
\label{sec:real_study}
\begin{figure}
\centering
\includegraphics[width=0.3\columnwidth]{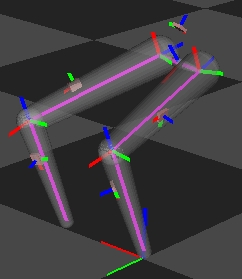}
\caption{Real data case study: biomechanical model with capsules.}
\label{fig:configurationsetup}
\end{figure}
In order to test the proposed method under real conditions, 
IMU data and global segment poses were captured at $240$ Hz from one subject ($33$ years, $1.85$m, $77$kg)
performing five squat exercises at normal speed
using the Xsens MVN BIOMECH Link inertial mocap system in lower body configuration (\cf \cite{Roetenberg2013}).
While the latter includes $7$ IMUs on feet, lower legs, upper legs and pelvis,
the feet IMUs were excluded from the study, since these were stationary during recording.
The segment lengths were measured and entered into the system manually and the N-pose option was used for calibration.
We also measured the leg circumferences manually in order to obtain the required radii 
for our proposed biomechanical model, which is illustrated in \figref{fig:configurationsetup}.
In summary, the setup included $5$ IMUs on right and left upper and lower legs and pelvis
$\mathbb{I} = \{I_{ru},I_{rl},I_{lu},I_{ll},I_p\}$, mounted on a biomechanical model with 
$5$ segments $\mathbb{S} = \{S_{ru},S_{rl},S_{lu},S_{ll},S_p\} 
(\|p^S_{ru/lu}\| = 0.546m,\|p^S_{rl/ll}\| = 0.440m)$.
Respective capsules $\mathbb{C} = \{C_{ru},C_{rl},C_{lu},C_{ll}\}$ 
($r_{p,ru/lu}=0.074m, r_{d,ru/lu}=r_{p,rl/ll}=0.049m, r_{d,rl/ll}=0.027m$) 
were only modelled for the legs and the $4$ \itos calibrations for the legs were in the focus of the analysis.
The hips were represented as ball-and-socket joints, while the knees were represented as hinge joints
with rotation axes obtained from \cite{Seel2014}:  
$\mathbb{J} = \{J_{lh},J_{rh},J_{lk},J_{rk}\}$, 
$\mathbb{H} = \{J_{lk},J_{rk}\}, h^S_{lk} = [0.159, -0.979,
0.124]^T, h^S_{rk} = [0.152, 0.951, 0.138]^T$, \rom $=[\theta_{min,lk/rk}=0,\theta_{max,lk/rk}=162]^\circ$.
The endpoints of the lower legs were assumed to have fixed positions during the squat movement.
The target \itos calibrations were chosen as indicated in \figref{fig:configurationsetup}.
In this study, a batch size $w=5$ was used to lower processing time.
For our case study, the \itos rotations and segment poses, as extracted from the captured data,
as well as, manually measured \itos positions (since these are not provided by the capturing system), 
were considered as ``well established'' reference values. 
Note, since these data cannot be considered ground truth (\cf Section \ref{sec:intro} and \cite{Zhang2013}), 
the focus of the real data case study was on repeatability of the calibration results.

Compared to the simulation study, the real data case study shows the following additional 
challenges for the proposed method:
\begin{itemize}
\item The capsule model (\cf \figref{fig:configurationsetup}) is only a rough approximation 
of the subject's body shape and the \itos poses therefore do not perfectly coincide with the shape prior.
\item The IMU data is noisy and biased (note, the gyroscope bias has been approximated from a stationary 
sequence and  subtracted in a preprocessing step).
\item The knee is not a perfect hinge and the axis is estimated.
\item Not all \dofs of the biomechanical model, in particular the hips, 
are fully excited during the squat motion.
\item All \itos calibrations for the leg IMUs are simultaneously initialised incorrectly.
\end{itemize}

The ability of the proposed method to produce repeatable calibration results
under these challenging conditions was tested by using three different \itos initialisation scenarios, namely:

(1) \textsl{plausible:} refers, for both legs, to the same initialisation as
chosen in the simulation study (\cf \figref{fig:twosegchain}).

(2) \textsl{simple:} refers to a default configuration, where all IMUs were initialised
at the middle of the segment on the capsule surface, their local $z$- and $y$-axes being aligned
with the segments' $x$- and $y$-axes, respectively. Note, this initial configuration is consistent with the shape prior.

(3) \textsl{perfect:} refers to the \itos orientations obtained from the reference system via the N-pose calibration
and the measured \itos positions.
\begin{figure}
\centering
\includegraphics[width=\columnwidth]{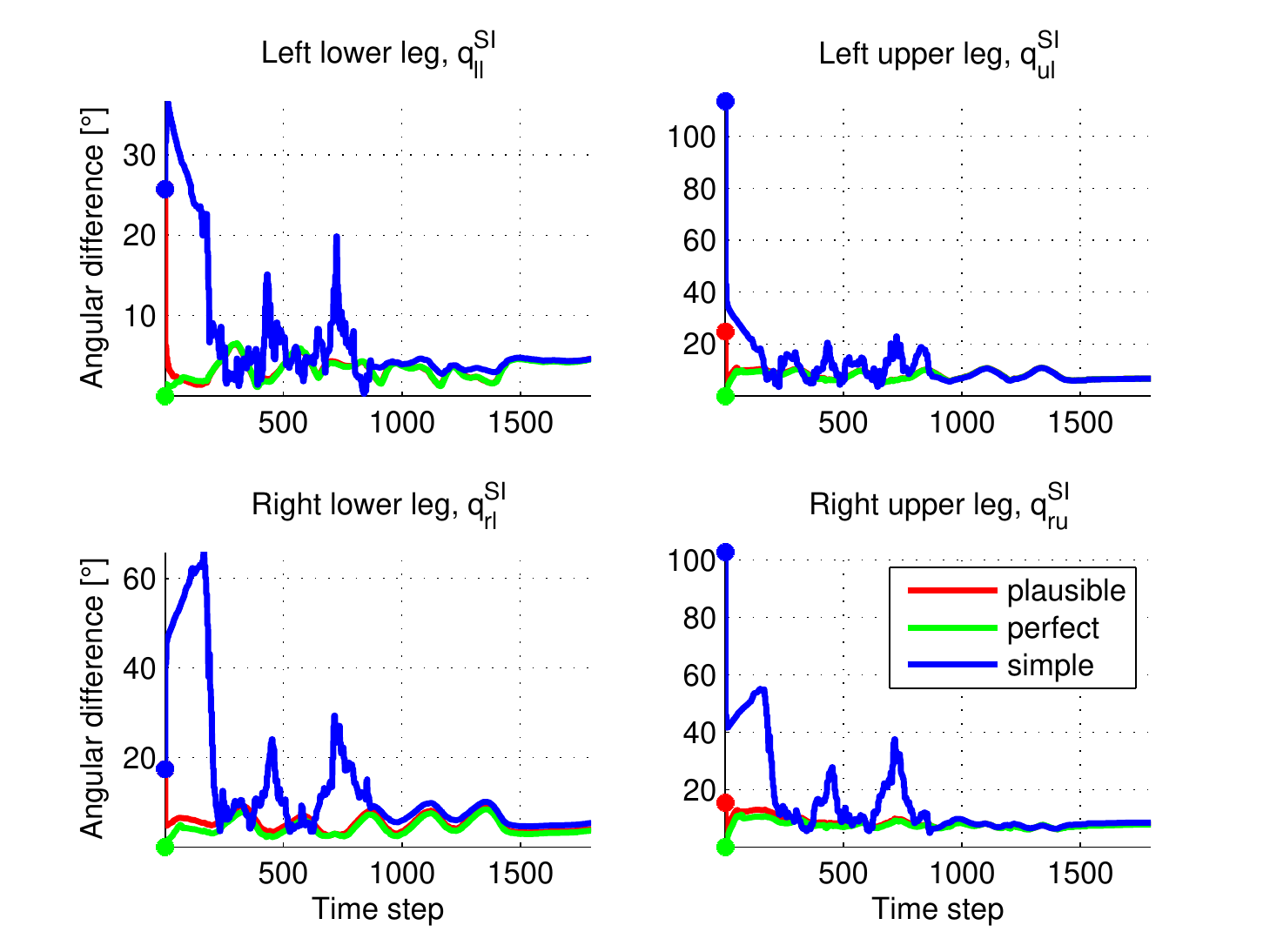}
\caption{Real data case study: convergence of the different \itos orientations (difference \wrt reference).}
\label{fig:RealDataI2S}
\end{figure}
\begin{table}
\caption{Real data case study: results for three different \itos initialisations.}
\centering
\begin{tabular}{|l|c|c|c|c|}
\hline
& \multicolumn{2}{|c|}{\textbf{Diff. to ref.: min, max}} & \multicolumn{2}{|c|}{\textbf{In-between diff.: min, max}} \\
\hline
 & \textbf{initial} & \textbf{final} & \textbf{initial} & \textbf{final} \\\hline
$\mathbf{{q^{SI}}_{ll}} [\circ]$  &$0.000,\, 25.73$  & $4.516,\, 4.613$ & $0.000,\, 25.73$ & $0.000,\, 1.178$ \\\hline
$\mathbf{{q^{SI}}_{lu}} [\circ]$  &$0.000,\, 113.6$  & $6.634,\, 6.664$ & $24.76,\, 113.6 $ & $0.079, \, 0.417$\\\hline
$\mathbf{{q^{SI}}_{rl}} [\circ]$ &$0.000,\, 17.36$  & $3.620,\, 5.295$ & $0.000,\, 17.36$ & $0.405,\, 1.707$\\\hline 
$\mathbf{{q^{SI}}_{ru}} [\circ]$ &$0.000,\, 102.8$  & $7.777,\, 8.451$ & $15.27,\, 102.8$& $0.371,\, 1.287$ \\\hline 
$\mathbf{{I^S}_{ll}} [m]$  &$0.000,\,0.076$  & $0.036,\, 0.066$ & $0.000,\, 0.077$ & $0.028, \, 0.062$ \\\hline
$\mathbf{{I^S}_{lu}} [m]$  &$0.000,\,0.093$  & $0.080,\,0.090$ & $0.055,\, 0.093$  & $0.004,\,0.012$\\\hline
$\mathbf{{I^S}_{rl}} [m]$  &$0.000,\, 0.077$  &$0.046 ,\, 0.060$ & $0.000,\, 0.077$ & $0.014 ,\, 0.030$ \\\hline
$\mathbf{{I^S}_{ru}} [m]$  &$0.000,\, 0.092$ & $0.061,\, 0.067$ & $0.055,\, 0.092$  & $0.003,\, 0.006$\\\hline
\end{tabular} 
\label{tab:diff-table}
\end{table}

Table \ref{tab:diff-table} summarizes the minimum and maximum differences of the
initial and final \itos calibrations of all test configurations \wrt the reference values, as well as,
among each other. Note, since the \textsl{perfect} initialisation is one of the test cases, 
the minimum initial differences to the reference are all zero. 
However, the maximal differences are above $113^\circ$ in orientation and $0.09m$ in position.
Looking at the final calibration results, these are all reasonably close to the reference calibration, 
with a maximum angular difference of $8.451^\circ$ and a maximum position difference of
$0.067m$ in the right upper leg. Recall, that the reference values cannot be considered ground truth.
We also confirmed that this difference was not introduced through the shape prior,
by observing that the same result was obtained when removing this prior and starting from the \textsl{perfect} calibration.

More importantly, all initial \itos calibrations converged 
to very similar final \itos orientations, with an in-between maximum angular difference of only 
$1.7^\circ$.
This is a promising result, which confirms the capability of the
algorithm to produce repeatable calibrations, as already indicated in the simulation study.

The maximum in-between \itos position difference was $0.062m$ for the right lower leg,
which is significantly larger than for the upper legs ($0.012m$, $0.006m$).
This might be explained by the low amount of motion in the lower legs as
compared to the upper legs during the squat exercise, yielding less information
for the calibration estimation.

\figref{fig:RealDataI2S} shows the evolutions of the \itos orientation
differences \wrt the reference for all IMUs. The figure clearly shows the different initialisations and the convergence to similar results
towards the end of the sequence. 
What can also be observed is a smooth but periodic change of all estimated \itos orientations during the sequence,
particularly in the calibration estimate of the lower right leg and the upper
left leg. This behaviour can have different sources, one of them being a time
dependent \itos calibration change due to soft-tissue artefacts. 
Such effects will be further investigated as part of our future work.

\section{Conclusion} \label{sec:conclusion}
This paper presents a method for simultaneous \itos calibration and 
body motion estimation from inertial sensors mounted on the body.
The method is based on sliding window constrained \wls optimisation
and combines state-of-the-art motion and measurement models with
different, partly novel biomechanical constraints, stochastic equations and priors.
Through experiments with simulated and real data, it has been shown that
the method can successfully estimate accurate and repeatable \itos calibrations from a wide range of initialisations.
For simulated data, the \itos calibrations converged reliably up to $95^\circ$ and convergences were observed up to a maximal tested initial angular offset of $131.19^\circ$, where the average precision was in the order of sub-degrees 
for the orientation and in the order of a centimetre for the position.
For real data, initialisations up to an initial angular difference of $113^\circ$ 
converged to similar results within a range of below $2^\circ$.

Given its online capable nature, the proposed method can not only be used for initial \itos calibration
without the need for precisely executed calibration poses or motions, but it could also be used 
for on-the-fly re-calibration, given an appropriate detection, \eg when an IMU slipped during recording.
This would significantly improve the usability of such systems.

\section*{Acknowledgment}
This work was performed by the junior research group wearHEALTH, funded by the BMBF (16SV7115).
For more information, please visit the website www.wearhealth.org.

\bibliographystyle{IEEEtran}
\bibliography{Fusion}

\end{document}